\documentclass[3p,twocolumn,authoryear]{elsarticle} 
\usepackage{graphicx}
\usepackage{amssymb}
\usepackage{hyperref}
\usepackage{bm}

\def\phs{\phantom{$-$}}

\def\mapiii{MAPPINGS {\sc iii}}

\newcommand{\hii}{H\,{\sc ii}}
\newcommand{\ha}{\ensuremath{\mbox{H}\alpha}}
\newcommand{\hb}{\ensuremath{\mbox{H}\beta}}

\newcommand{\ciii}{C\,{\sc iii}}

\newcommand{\nii}{N\,{\sc ii}}

\newcommand{\oii}{O\,{\sc ii}}
\newcommand{\oiii}{O\,{\sc iii}}

\newcommand{\neii}{Ne\,{\sc ii}}
\newcommand{\neiii}{Ne\,{\sc iii}}

\newcommand{\sii}{S\,{\sc ii}}
\newcommand{\siii}{S\,{\sc iii}}
\newcommand{\siv}{S\,{\sc iv}}

\catcode`\@=11
\def\gapprox{\mathrel{\mathpalette\@versim>}}
\def\lapprox{\mathrel{\mathpalette\@versim<}}
\def\@versim#1#2{\lower2.45pt\vbox{\baselineskip0pt\lineskip0.9pt
      \ialign{$\m@th#1\hfil##\hfil$\crcr#2\crcr\sim\crcr}}}
\catcode`\@=12
 
\newcommand{\pccm}{\ensuremath{\,\mbox{cm}^{-3}}}

\newcommand{\cms}{\ensuremath{\,\mbox{cm}\,\mbox{s}^{-1}}}
\newcommand{\kms}{\ensuremath{\,\mbox{km}\,\mbox{s}^{-1}}}

\newcommand{\mum}{\ensuremath{\mu\mbox{m}}}
\newcommand{\msun}{\ensuremath{\mbox{M}_{\odot}}}

\begin{document} 
\title{ITERA: IDL Tool for Emission-line Ratio Analysis}

\begin{frontmatter}
\author[strw]{Brent A. Groves}
\ead{brent@strw.leidenuniv.nl} 

\author[stras]{Mark G. Allen }
\ead{allen@astro.u-strasbg.fr}

\address[strw]{Sterrewacht Leiden, Leiden University, Neils Bohrweg 2, Leiden 2333-CA
       The Netherlands}
\address[stras]{Observatoire de Strasbourg UMR 7550 
       Strasbourg 67000, France}

\begin{abstract}
We present a new software tool to enable astronomers to easily compare
observations of emission line ratios with those determined by
photoionization and shock models, ITERA, the IDL Tool for
Emission-line Ratio Analysis.  This tool can 
plot ratios of emission lines predicted by models and allows for
comparison of 
observed line ratios against grids of these models selected from
model libraries associated with the tool. We provide details of the libraries of standard
photoionization and shock models available with ITERA, and, in
addition, present three example emission line ratio diagrams covering
a range of wavelengths to demonstrate the capabilities of ITERA.
ITERA, and associated libraries,  is available from
\url{http://www.brentgroves.net/itera.html}. 

\end{abstract}


\end{frontmatter}

\section{Introduction}

All emission lines are
sensitive to some extent on the physical conditions in the emitting
gas. This includes the density and electron temperature of
the medium, the abundance of the emitting element, and the details of the
ionizing source such as the luminosity and effective temperature of the star in \hii\
regions. While both atomic and line theory and observational 
data have advanced sufficiently over
the years such that the dependency of individual emission lines on
nebulae parameters is known, determining these same parameters from
individual lines is difficult apart from a few exceptions.  One of the
simplest ways to break these inherent degeneracies is through the use of
emission line ratios. Using lines arising from the same elements, or
similar ionization potentials, or even from different levels of the
same ion, can minimize
the dependence on some or most of the parameters that control line
emission, thus enabling the construction of diagnostics.

One of the first diagnostic emission line ratios was the
[\oiii] line ratio $\lambda 4363$\AA/$\lambda 5007$\AA\,  put forward by
\citet{Menzel41}. By using forbidden lines of the same ion, O$^{+2}$,
both the
abundance and ionization dependencies are removed, leaving a ratio strongly
dependent on electron temperature due to the different excitation
levels of the lines involved \citep[see e.g.~Fig.~2.5
in][]{Dopita03}. This line method, and other methods of determining
electron temperature, have been
discussed in detail by \citet{Peimbert67}. The same reasoning has also
been
extended to several other line ratios, using lines from a single 
ion but arising from different energy levels, such as optical-IR
ratios like [\oiii]$\lambda 5007$\AA/$\lambda 52\mu$m \citep[see e.g.][]{Dinerstein85} and UV ratios
like \ciii]$\lambda 1909$/\ciii$\lambda 977$.

Similarly, ratios of forbidden lines arising from the same ion and
similar energy levels remove abundance, ionization and temperature
dependencies. If these lines arise from levels with different
collisional de-excitation rates or radiative transition probabilities,
this ratio will be sensitive to the densities in the range spanned by the critical densities of
the lines involved. This idea was first put forward by \citet{Aller49} for the
[\oii] doublet, $\lambda3726/\lambda3729$, and has been extended to
several other strong lines like [\sii] $\lambda6716/\lambda6731$ or in
the IR [\siii] $\lambda18.7\mum/\lambda33.5\mum$ \citep[see][for a
nice overview of these ratios]{Rubin89}.

Apart from the temperature and density sensitive ratios described
above, determining physical quantities from
single line ratios proves difficult due to degeneracies between the
various controlling parameters of the nebula emission. 
Generally what is done is to use
several line ratios in combination to break these degeneracies, 
such as determining the
metallicity or abundances in the gas \citep[see e.g.][for an overview of
several line methods for metallicity determination]{Kewley08}, or to
use empirically-based assumptions, such as the constancy of the Balmer
decrement in \hii\ regions for dust-reddening determination
\citep[also theoretically supported, see e.g.][ though these concentrate
mostly on planetary nebulae]{Cox69, Brocklehurst71,Miller72}. An
overview of all methods of comparing observed emission lines with
theory can be found in \citet{Stasinska07} and in books such as \citet{Osterbrock06} and
\citet{Dopita03}, which give a broad understanding of the topic and
underlying physics.

One of the best ways to disentangle these degeneracies is to plot two
line ratios against each other in what is commonly called a line
diagnostic diagram. These were first used as a method to
distinguish various classes of emission-line galaxies in
\citet{Heckman80} and \citet{Baldwin81}, with the diagrams described
in the latter still commonly used to distinguish star-forming or \hii\
galaxies from galaxies dominated by an active galactic nucleus
(AGN). These diagrams for emission-line galaxy classification 
were revised and extended by \citet{Veilleux87},
who based their selection of line ratios on 5 criteria: Line strength (1),
Line proximity (for blending (2) and reddening (3) effects), the preferable
inclusion of hydrogen lines (4) and the capability of detecting these
lines (5). 

With improvements in the availability of spectra, with wavelengths from
the UV (such as Hubble-STIS) to FIR (e.g.~Herschel -PACS \& -HIFI)
available, increased spectral sensitivity and resolution, and a
copious number of ionization models now
available \citep[e.g.][]{Morisset08}, such criteria need not be so
strict.  However, these large increases in wavelength range,
sensitivity, and models make analysis of emission-line objects much
more complicated.

It is for these reasons that we introduce here a new, publicly available, emission-line
analysis tool, ITERA. ITERA, the IDL Tool for Emission-line Ratio
Analysis, allows the simple comparison of observed emission-lines from
UV to far-IR wavelengths to models to assist in the determination of
properties such as gas density, metallicity and excitation mechanism. 
We describe in the following sections the code and demonstrate some of
the possible diagnostic diagrams available.

\section{The ITERA Program}

The IDL Tool for Emission-line Ratio Analysis (ITERA) had its
inception with the SHOCKPLOT tool created to display the results of 
the \citet{Allen08} 150--1000\kms\ shock models. The power of such a
tool was quickly realised and the original tool was rewritten and
expanded into its current form. ITERA is an IDL\footnote{IDL, the
  Interactive Data Language, is a computing environment for data analysis, 
visualization, and application development, available from ITT Visual
Information Solutions (\url{http://www.ittvis.com/ProductServices/IDL.aspx}).}
widget tool which enables astronomers to plot line ratio 
diagrams of emission lines arising from atomic and ionized species as determined
by standard photoionization and shock models. It also allows for a
comparison of these line ratios with those obtained from
observations or other models. As discussed in the
introduction, these diagrams, or the individual ratios of the diagram,
can be used as diagnostics of nebulae excitation mechanisms, nebulae
density, electron temperature, metallicity or individual abundance
variations, and other model-specific parameters. ITERA can also be
used to determine line sensitivities to such parameters, thus giving
the possibility of creating new or ``best possible''  diagnostics for
a given set of observations. It can even be used to estimate line
fluxes that are weak or in unobserved wavelength ranges.

ITERA works in three simple steps; the user selects the emission lines
to be used as the
denominators and numerators of the ratios for the X- and Y- axes,
selects the model(s) that the user wants to examine, and, if wished,
the user enters observational data for comparison.  These steps can be
performed in any order. Multiple lines can be chosen for the ratios in
the case of emission line doublets or line blends due to
resolution or velocity effects.
What is returned is
an emission-line ratio diagram of the chosen ratios, displaying a grid
of the chosen models and, if entered, observational data points. This
diagram can be printed, or the model grids displayed can be output for
use in other routines, or direct comparison with data.
The program and associated libraries and routines are available from
\url{http://www.brentgroves.net/itera.html}. An IDL virtual
machine\footnote{IDL virtual machine is free runtime version of IDL
available from 
\url{http://www.ittvis.com/ProductServices/IDL/VirtualMachine.aspx}.} 
version is also available for those without access to IDL licences. 
Full instructions and details of the program are found on the webpage,
but we describe below the major parts of the program.   

The principle of ITERA is to allow
ease of access for astronomers to existing shock- and photo-ionization
models for the determination of emission line strengths. Thus the
strength of ITERA lies in the 
library of shock- and photo- ionization models associated with the
code. This library is a distinct yet necessary part of ITERA, needed
to generate the line ratios. As the library is not an integral
part of ITERA, only the desired model series need to be downloaded and
installed (such as AGN or  \hii-region models, with further models
added later as required, including photoionization models generated by
the user.  

\subsection{Choosing the Models}

The library of models associated with ITERA cover the range of
excitation mechanisms expected within emission-line galaxies; star
formation, active galactic nuclei and shocks. These models, generated
by the shock and photoionization code \mapiii, broadly cover the ratio space
occupied by the dominant optical emission lines from galaxies. Other
sources of emission lines, such as individual O and B stars (\hii\
regions) or planetary nebulae, can be included through the user-generated
option of ITERA, and may be included in the library at a later date.
The ITERA model library is split into four main categories: 
Starburst or stellar ionization models, AGN models, shock
models, and user-generated models. The currently available libraries
are shown in table \ref{tab:models}, and described briefly below.

ITERA has the possibility of plotting up to any number of shock or
photoionization model grids, although for clarity the number displayed on any
plot is generally limited to 5. Every model set has a limited number of
parameters, with common parameters to the models being the emitting gas
metallicity or abundance pattern, and the gas density or pressure. 
The other parameters associated with the models tend to be specific
to the specific sets, either due to the different physics
associated with the ionizing/excitation mechanism, or due to different
definitions by the model set creators. These include the hardness of the radiation
field, the velocity of a shock ionizing the gas, or the
dimensionless photoionization parameter. The parameters associated
with the different model sets within the ITERA model library are
described in the following subsections.  

The model grids displayed within the ITERA plot window are formed by
the ranges of two of these parameters, with the other parameters of the model
sets fixed. The fixed parameters thus describe the different model
grids that can be plotted. The first parameter of the model grid is
set and cannot be chosen by the user, and is generally a measure of
the ionization state of the gas, such as ionization parameter or shock
velocity. The second parameter of the grid can be chosen by the user,
with the choice of density, metallicity or the unique parameter
associated with that model set (as described below). If the user of
ITERA wants to compare parameters other than the set first parameter,
e.g~the ionization parameter, this can be done by simply
limiting the parameter range to a single value and plotting multiple grids.


\begin{table}[htp]
\begin{tabular}{|l|}
\hline
{\bf Starburst models} \\
\hline
\citet{Kewley01} models\\
~$\bullet$ Pegase2 stellar models\\
~~$\circ$ Instantaneous Star Formation \\
~~$\circ$ Continuous Star Formation \\
~$\bullet$ Starbust99 stellar models\\
~~$\circ$ Instantaneous Star Formation \\
~~$\circ$ Continuous Star Formation \\
\citet{Levesque10} models \\
~$\bullet$ Standard mass loss stellar models\\
~~$\circ$ Instantaneous Star Formation \\
~~$\circ$ Continuous Star Formation \\
~$\bullet$ High mass loss stellar models\\
~~$\circ$ Instantaneous Star Formation \\
~~$\circ$ Continuous Star Formation \\
\citet{Dopita06} ${\cal R}$ parameter models\\
\hline
{\bf AGN models} \\
 \hline
Standard, constant density, dust-free, \\
~~photoionized AGN \\
Dusty, Radiation-Pressure dominated, \\
~~photoionized AGN \\
\hline
{\bf Shock Models} \\ 
\hline
Shock ionized cloud \\
Photoionized Precursor \\
Shock$+$Precursor\\
\hline
{\bf User-generated models}\\
\hline
User run Cloudy \& \mapiii \\
~~photoionization models\\
\hline
\end{tabular}
\caption{Listing of the model libraries currently available with ITERA}\label{tab:models}
\end{table}

\subsubsection{Starburst Models}

The ITERA model library includes 3 different Starburst models, aimed
specifically at galaxies whose line emission is dominated by gas
ionized by recent ($< 10^7$ yr) star formation, generally classified
as starburst, \hii, or star-forming galaxies.  However, these models
can also be
generally applied to any massive star formation region, though the
comparison may break down for less massive star formation regions once
the effects of the stochastic sampling of the stellar initial mass function
becomes important ($\lapprox 10^4$ \msun).

The first model set is from \citet{Kewley01}. These models use
stellar ionizing spectra generated by the stellar population synthesis
codes PEGASE-2 \citep{Fioc97} and Starburst99 \citep{Leitherer99},
and considers both continuous star formation and instantaneous
bursts. These models are then passed through the \mapiii\ code for a
range of ionization parameters\footnote{The ionization parameter, $q$,
  also known as the ionization front speed, is a
  measure of the ionizing photon flux over the (hydrogen) gas
  density; $q=S_{*}/n_{\rm{H}}$, where $S_{*}$ is the integral of the
  photon flux above 13.6 eV. This is also often given in a
  dimensionless form; ${\cal U}=q/c$.}. The stellar models and subsequent
photoionization modelling are determined for 5 metallicities, ranging
from 0.05 to 2.0 Solar ($Z_{\odot}$). For full details see the
\citet{Kewley01} paper. These models were used as a basis for the strong-line
metallicity estimators in \citet{Kewley02} and for the
empirical/theoretical AGN/star-formation dividing lines discussed in
\citet{Kauffmann03} and \citet{Kewley06}. 

The second set of models is from \citet{Levesque10}, and is an updated
version of the \citet{Kewley01} models, using a later version of
\mapiii\ (version {\sc iii}r) code for the radiative transfer, and
using the latest version of the Starburst99 code \citep{Vazquez05}. Like
the \citet{Kewley01}  work, \citet{Levesque10} model includes both
instantaneous burst and continuous star formation stellar models, 
and covers a similar range in ionization parameter, $q$, and
metallicity. In addition, the \citet{Levesque10} models also
cover a range of ages between 0 to 10 Myr (in steps of 0.5 Myr). What
distinguishes these models from previous works is that they examined two
different evolutionary tracks for the stellar populations, one with a
standard mass loss treatment (STD) and one with high mass loss rates
(HIGH) meant to more accurately match observations \citep[see][for
full details about the different tracks and the models]{Levesque10}.

The third set is the ``${\cal R}$ parameter'' models of
\citet{Dopita06}. As with the previous set, these use \mapiii r for
the radiative transfer and the latest version of Starburst99 to
generate the ionizing stellar spectra, though only instantaneous
bursts were considered. The same range of stellar ages and
metallicities as \citet{Levesque10} are used. What differentiates
these models from the previous sets is that a direct link with the
stellar populations and the surrounding ionized gas is
considered. In these models all stellar clusters are born buried within
their molecular birth clouds, but, over time, they clear an empty
bubble around them, sweeping up the gas that forms the surrounding
\hii\ regions. This means that there is a direct link of both the gas
density and incident ionizing flux, and thus the ionization parameter,
at the inner edge of the \hii\ region to the evolution of the
stars. With this link, the time-averaged ionization parameter is found
to be directly correlated with the mass of the ionizing cluster
divided by
the average ISM pressure, defined within the \citet{Dopita06} work as ${\cal R}$
($=\rm{M}_{\rm{cl}}/P_{\rm{ISM}}$). Thus it is this
${\cal R}$ parameter rather than the local ionization parameter, $q$,
that is measured when the emission-line ratios are measured on
galactic scales. For full details about the ${\cal R}$ parameter and
the models see \citet{Dopita06}.

One important caveat with these models is that, while all three use
the same global metallicities, the elemental abundances at each
metallicity for each model differ due to the evolving definition of
the abundance pattern at solar metallicity. Thus such differences must be
considered when comparing these models. In table \ref{tab:metals} we
list the solar abundance patterns used in all models of the ITERA
library discussed here. Most abundances are the same, except for
oxygen which was recalibrated between different model sets \citep[see
e.g.][]{Asplund05}. Different metallicities simply scale as multiples
of the abundances listed here, except  for helium and nitrogen, a secondary element
\citep[see][]{Groves04a}. One cautionary note on the abundances listed
here is that these do not take account of the effects of dust
depletion, which varies amongst the models (from non-existent to
significant), which will alter the gas-phase abundances.

\onecolumn
\begin{table}[htp]
\begin{tabular}{|lccccc|}
\hline
Element  & Kewley2000 & Levesque2010 & ${\cal R}$ param & AGN models & Shocks \\
\hline
H..........& \phs0.00 & \phs1.00 & \phs0.00 & \phs0.00 & \phs0.00  \\
He.........& \phs1.01 &  -1.01   & -1.01 &  -0.99 &  -1.01 \\ 
C..........& -3.44    &  -3.59   & -3.59 &  -3.61 &  -3.44 \\ 
N..........& -3.95    &  -4.22   & -4.22 &  -4.20 &  -3.95 \\ 
O..........& -3.07    &  -3.34   & -3.34 &  -3.31 &  -3.07 \\ 
Ne.........& -3.91    &  -3.91   & -3.91 &  -3.92 &  -3.91 \\   
Na.........&       ~  &  -5.75   &     ~ &  -5.68 &      ~ \\ 
Mg.........& -4.42    &  -4.47   & -4.47 &  -4.42 &  -4.42 \\ 
Al.........&       ~  &  -5.61   &     ~ &  -5.51 &  -5.53 \\ 
Si.........& -4.45    &  -4.49   & -4.49 &  -4.49 &  -4.45 \\ 
S..........& -4.79    &  -4.79   & -4.79 &  -4.80 &  -4.79 \\ 
Cl.........&       ~  &  -6.40   &     ~ &  -6.72 &      ~ \\ 
Ar.........& -5.44    &  -5.20   & -5.20 &  -5.60 &  -5.44 \\   
Ca.........& -5.64    &  -5.64   & -5.64 &  -5.65 &  -5.88 \\   
Fe.........& -4.33    &  -4.55   & -4.55 &  -4.54 &  -4.63 \\   
Ni.........&       ~  &  -5.68   &     ~ &  -5.75 &      ~ \\ 
\hline
\end{tabular}
\caption{Solar abundances used in ITERA Libraries, given as log$_{10}$(X/H).}\label{tab:metals}
\end{table}
\twocolumn

\subsubsection{AGN Models}

The library of AGN models used in the ITERA code are those from the
\citet{Groves04a,Groves04b} photoionization models of the narrow line
regions of AGN.  We include two series of AGN models, the classic
dust-free, AGN power-law ionized models and the dusty, radiation
pressure dominated models introduced by Groves et al.. Both of these
models use simple power-law radiation fields
($f_{\nu}\propto\nu^{\beta}$) to represent the AGN ionizing spectrum,
and explore the same range in metallicities (with the same abundance
patterns), gas densities and ionization parameters. However the first
models are dust free and assume a constant gas density throughout the
narrow line region cloud. The second set are taken to be isobaric
(constant pressure) and include dust and the effects of radiation
pressure on dust, which causes significant differences in the
resulting emission line spectrum at high ionization parameters
\citep[see][for full details of the
models]{Groves04a,Groves04b}. These latter models are much more
closely matched with the observed line ratio space and are the
recommended model to use.

While these models do not represent the full range of theoretical
models for the emission from the narrow line regions of AGN \citep[see
e.g.,][]{Binette96, Ferguson97}, they do broadly cover the expected
emission of strong emission lines, and can be used to estimate the ``AGN
space'' of line ratios.

\subsubsection{Shock Models}

The shock models used in the ITERA library are directly imported from the
\citet{Allen08} work on fast shocks. They include a range of metallicities,
densities, and magnetic parameters (the ``unique parameter'' of these
models), and explore velocities from $\sim 
150$ to $1000\kms$, where strong J-shocks occur. The shock models are
split into three sets. The first is the emission from the post-shock
region, where gas is shock-excited to high temperatures and ionization
states. The second model set is from the pre-shock or precursor region
which is photoionized by the radiation emitted upstream from the
post-shock region. The third set is the combination of these
(shock+precursor) for integrated measurements where the individual
regions cannot be distinguished. For full details on the parameters
and underlying physics of the models see \citet{Allen08}.

\subsubsection{User Generated Models}

While the above models cover the expected parameter space of
emission-line galaxies, they are not representative of individual
objects, where variations in the abundances of specific elements,
ionizing source variations or even geometrical effects can all affect
the final emission-line spectrum. Therefore we include the final
option of user generated model libraries. This allows the possibility
of a user inputting the results from well tested and
respected photoionization codes such as Cloudy\footnote{Cloudy is
  available for download from \url{http://www.nublado.org}, along with
full details on the code, how to use it, and how it works.}
\citep{Ferland98} and \mapiii\footnote{\mapiii\ is available to
  download from \url{http://www.brentgroves.net/mapiii.html},
  which includes a description of the code and its use.}
\citep{Groves04a}, or using on-line interfaces such as mapiiionline:
\url{http://www.ifa.hawaii.edu/~kewley/Mappings/}. For a listing,
comparison, and benchmarking of such models see \citet{Pequignot01},
though all these models have advanced since this work.

To assist in the input of user generated models into ITERA, we provide 
tools for converting the outputs from these models to format
used by ITERA. Currently only conversion tools for the publicly available codes
Cloudy and \mapiii\ are provided, but this may be expanded at a later
date depending on demand. 

\subsection{Choosing Line Ratios}

In principle, the choice of emission lines to use in the diagnostic line
ratios in ITERA is limited purely by those provided by the code used to
generate the photoionization/shock model. In \mapiii\ (including all
the models provided in the library) this is limited to $\sim$1770
emission lines in the infrared, optical, and ultraviolet, and
even including high-ionization/inner-shell lines in the X-ray. In
Cloudy, this is further expanded by a host of weaker lines as well as
molecular ro-vibrational emission lines from CO  and H$_2$. The lines
are listed in simple selection boxes, with only strong lines listed by
default, though all lines, or lines in certain wavelength ranges, can
be displayed if wished.  

In addition, sums of lines can be used in the line ratios to account
for line blending at low spectral resolution, or to account for
different levels or ionization states (i.e.~[\oii]$\lambda 3726+3729$). 

In practice however, the guidelines discussed in the introduction form
the bases for good diagnostics. Lines of the same element and of the same
ionization state or of similar excitation potential in different
elements provide some of the strongest diagnostics. Lines relative to
hydrogen recombination lines are also good diagnostics. As discussed
in \citet{Veilleux87}, lines in close wavelength proximity limit the
possible instrumental, calibration, and reddening effects.

Hence it is for this reason that several standard, strong-line,
diagnostic ratios, such as [\sii]$\lambda\lambda6716,31/\ha$ versus
[\oiii]$\lambda5007/\hb$, are also 
included in ITERA as predefined diagnostic diagrams for ease of use. 

\subsection{Comparison with Observations}

Ultimately, the use of ionization models lies in their diagnostic
power for observations, which is why ITERA also includes the ability
to enter data.
To allow for easy comparison of observations with the models, the
ITERA program has three simple ways in which data can be entered into the
emission-line diagnostic diagrams directly; direct point by point
entry, a simple list of line ratios for multiple data, and via tabulated emission line
data for multiple  line ratio comparisons. For the first two options,
the line ratios for the chosen diagram are entered in and compared
directly with the model on the chosen line ratio diagram (as shown in
figure \ref{fig:midIR}). Once the chosen diagram has cleared or another
line ratio diagram chosen, the data are cleared. The tabulated
emission line data are chosen in a similar manner as the
models, with the data plotted over the model grids if  the contain the
chosen emission lines. Templates for the tabulated data are available
with the ITERA program.

Conversely, the fluxes (relative to H$\beta$) from the selected model
grids for the four emission lines of the chosen line diagnostic
diagram can be exported to formatted ASCII files for comparison with
observational data. This is especially useful in cases of large
datasets, such as with SDSS.

\section{Emission Line Ratio Diagnostics}

To demonstrate the capabilities of ITERA, we present here three
emission-line ratio diagrams covering a range of wavelengths and
diagnostic capabilities. These diagrams are all direct output from the ITERA
routine, excluding the histogram in the first diagram.

\begin{figure}
\includegraphics[width=\hsize]{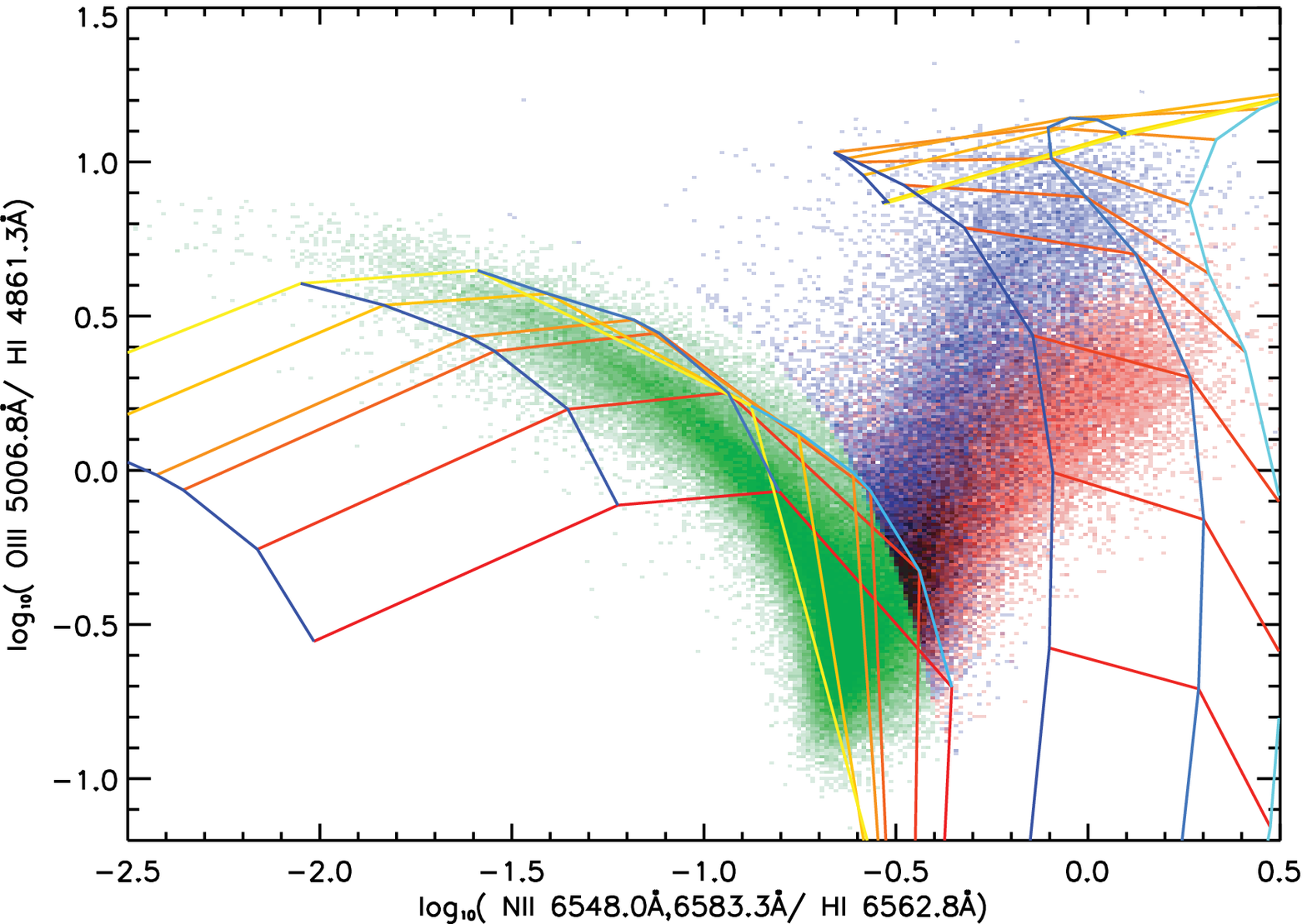}
\caption{The BPT diagram of [\nii]/\ha\
versus~[\oiii]/\hb, showing grids of photoionization models for star
forming galaxies \citep[left,][]{Levesque10} and AGN
\citep[right,][]{Groves04b}, described fully in the text. In both grids, lines of constant
metallicity are shown by the blue curves, increasing from dark blue to
light blue, and lines of constant ionization parameter are shown by
the orange curves, increasing in value from red to yellow. In the
background is an image of a 2d density diagram from the SDSS DR4
sample described in \citet{Groves06}, showing star-forming galaxies
(green), Seyfert galaxies (blue) and LINER galaxies (red).}
\label{fig:BPT}
\end{figure}

The first diagram (figure \ref{fig:BPT}) shows [\nii]/\ha\
versus~[\oiii]/\hb, the classic ``BPT'' diagram from \citet{Baldwin81}
that is a diagnostic of the excitation mechanism in emission-line
galaxies.  Shown on this plot are two model grids; the
\citet{Levesque10} STD models for star-forming galaxies on the left, and the
\citet{Groves04b} models for Dusty AGN on the right. Both grids show a range of
metallicities (blue curves) and ionization parameters (orange curves),
though note that these grids are not equivalent. The exact parameters that
the star forming models have are:
$Z\sim0.05,0.2,0.4,1.0,2.0Z_{\odot}$ and $q=1\times10^7, 2\times10^7,
4\times10^7, 8\times10^7, 1\times10^8, 2\times10^8, 4\times10^8\cms$,
for a 2 Myr old burst with density $n_{\rm H}=100\pccm$,
while the AGN models have: $Z\sim1.0,2.0, 4.0Z_{\odot}$ with
$\log {\cal U}=0.0,-0.3,-0.6,-1.0,...,-3.6,-4.0$, a gas density of
$n_{\rm H}=1000\pccm$ and are ionized by a power-law of index
$\beta=-1.4$. Underlying the grids is a coloured 2D density histogram
of the SDSS DR4 emission line galaxy sample from \citet{Groves06}, displaying
the observed distribution of star forming galaxies (Green squares), Seyfert
galaxies (blue squares), and LINER galaxies (red squares).  

The model grids clearly encompass the observed line ratio space. 
Note how the star forming models follow closely the green branch of
this diagram, showing the known strong dependence on metallicity of
the [\nii]/\ha\ ratio, with the ionization parameter affecting the
position of the models in both ratios, explaining the spread in
observed galaxies (along with the average age of the ionizing
stars). Also note that the models fail to reproduce the observed line
ratios at very low metallicities, and
do so even when different ages are used for the ionizing stars,
indicating that there are still underlying issues with either the photoionization
models, stellar population models, or both.
The AGN model grid on the other hand actually tends to be too broad,
demonstrating the sparsity of low metallicity AGN in the SDSS
emission-line galaxy sample, as discussed in \citet{Groves06}. The low
${\cal U}$ NLR models are likely to be too faint to be observed in this sample.
It should be remembered that the AGN and star forming galaxy models
show are the extremes, of  
pure AGN contribution and pure \hii\ regions. In between these models,
seen in the right-hand wing  
of the diagram (the `AGN' branch) are galaxies where both mechanisms, active star
formation and AGN, contribute to the galaxy spectrum.

\begin{figure}
\includegraphics[width=\hsize]{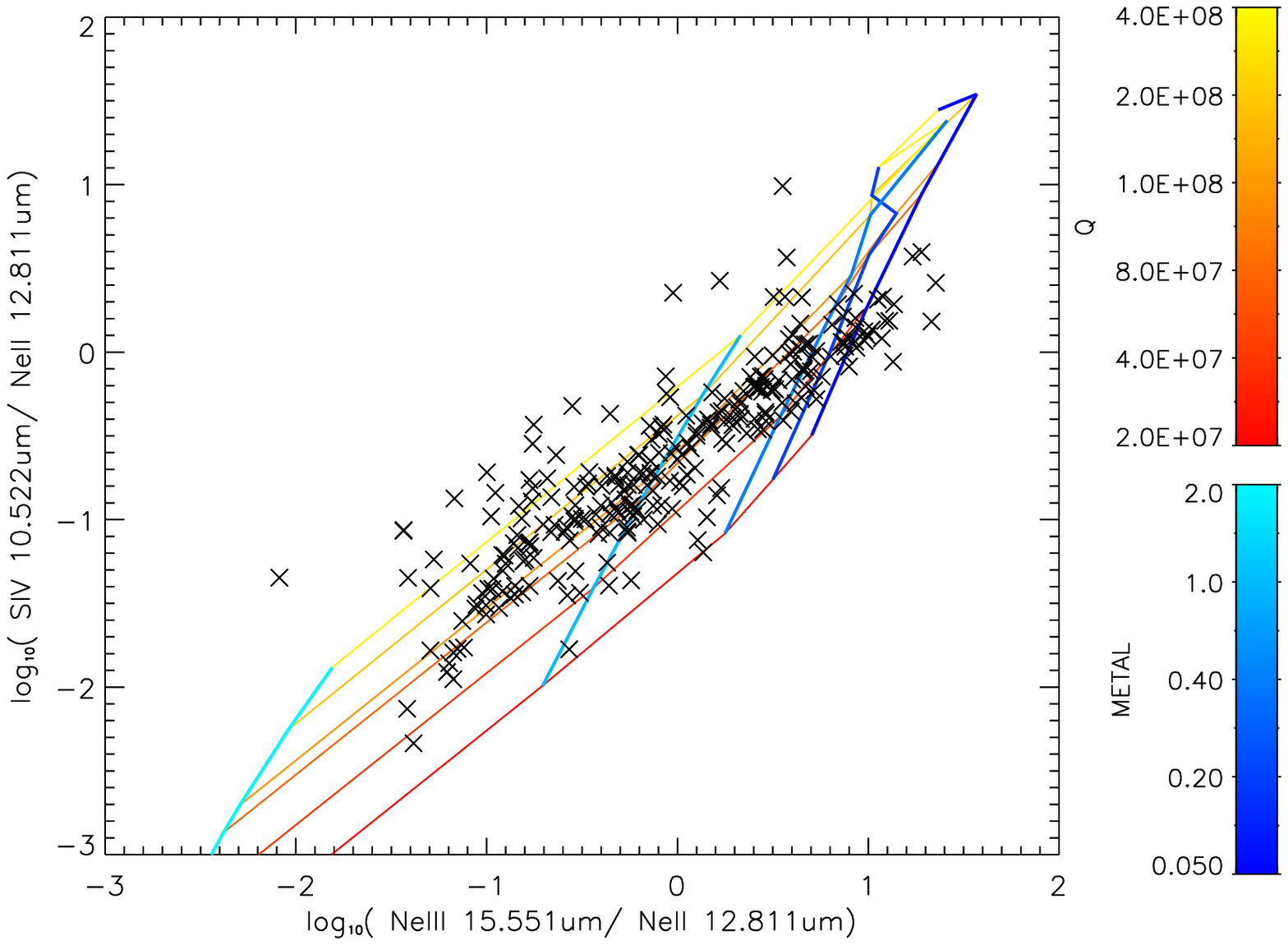}
\caption{The mid IR emission line ratio diagram of
  [\neiii]${15.6\mu\rm{m}}$/[\neii]${12.8\mu\rm{m}}$ versus 
[\siv]${10.5\mu\rm{m}}$/[\neii]${12.8\mu\rm{m}}$, showing the same
\citet{Levesque10} star forming galaxy photoionization model as in
figure\ref{fig:BPT}, with the grids curves as labelled in the colour
bar to the right. The stars are a sample of O and B star
ionized systems (star-forming galaxies and extragalactic and galactic
\hii\ systems) collated and described in \citet{Groves08}.}
\label{fig:midIR}
\end{figure}

The second diagram (figure \ref{fig:midIR}) shows the mid-IR diagram of
[\neiii]${15.6\mu\rm{m}}$/[\neii]${12.8\mu\rm{m}}$ versus
[\siv]${10.5\mu\rm{m}}$/[\neii]${12.8\mu\rm{m}}$. A surprisingly
strong correlation between these ratios was found by \citet{Groves08},
who explored a large sample of emission line objects to see if the
ground-observable [\siv]/[\neii] ratio could replace the space-only,
ionization sensitive [\neiii]/[\neii] ratio. Here we show only a sub-sample
of that collated by \citet{Groves08}, displaying only star formation
ionized regions (starbursts, blue compact dwarfs, extragalactic \hii\
regions, and galactic \hii\ regions), shown as crosses on figure
\ref{fig:midIR}. Overlaying the observed data is same star-formation
model grid of metallicity and ionization parameter as in figure
\ref{fig:BPT}, as labelled in the colour bars. The model grid broadly
covers the observations, with both higher metallicity and lower
ionization parameter towards the  lower left of the grid (lower
ionization). This is somewhat degenerate with the age of the ionizing
stellar population, with ages greater than $\sim4$ Myrs moving
strongly to the lower left of the grid. 

There are two interesting
points to take from this grid. The first is that the models are unable
to explain the full spread of the data, though this may arise due to
the $\sim 20\%$ uncertainty in the data \citep[see][for
details]{Groves08}. The high values of [\siv]/[\neii] may come also
from higher ionization parameters and or different abundance patterns,
but the points to the right of the diagram suggest, as in the previous
figure, that the low metallicity models (or youngest age or Wolf-Rayet
models similarly) are lacking a hard enough spectrum to reach these
ionization states.

The second is that the observed relation between [\neiii]/[\neii]
and [\siv]/[\neii] has a different slope and narrower spread than the
model grid, suggesting that there may a higher order correlation going
on between the metallicity and age of the stellar population and the
ionization parameter of the ionized gas.

\begin{figure}
\includegraphics[width=\hsize]{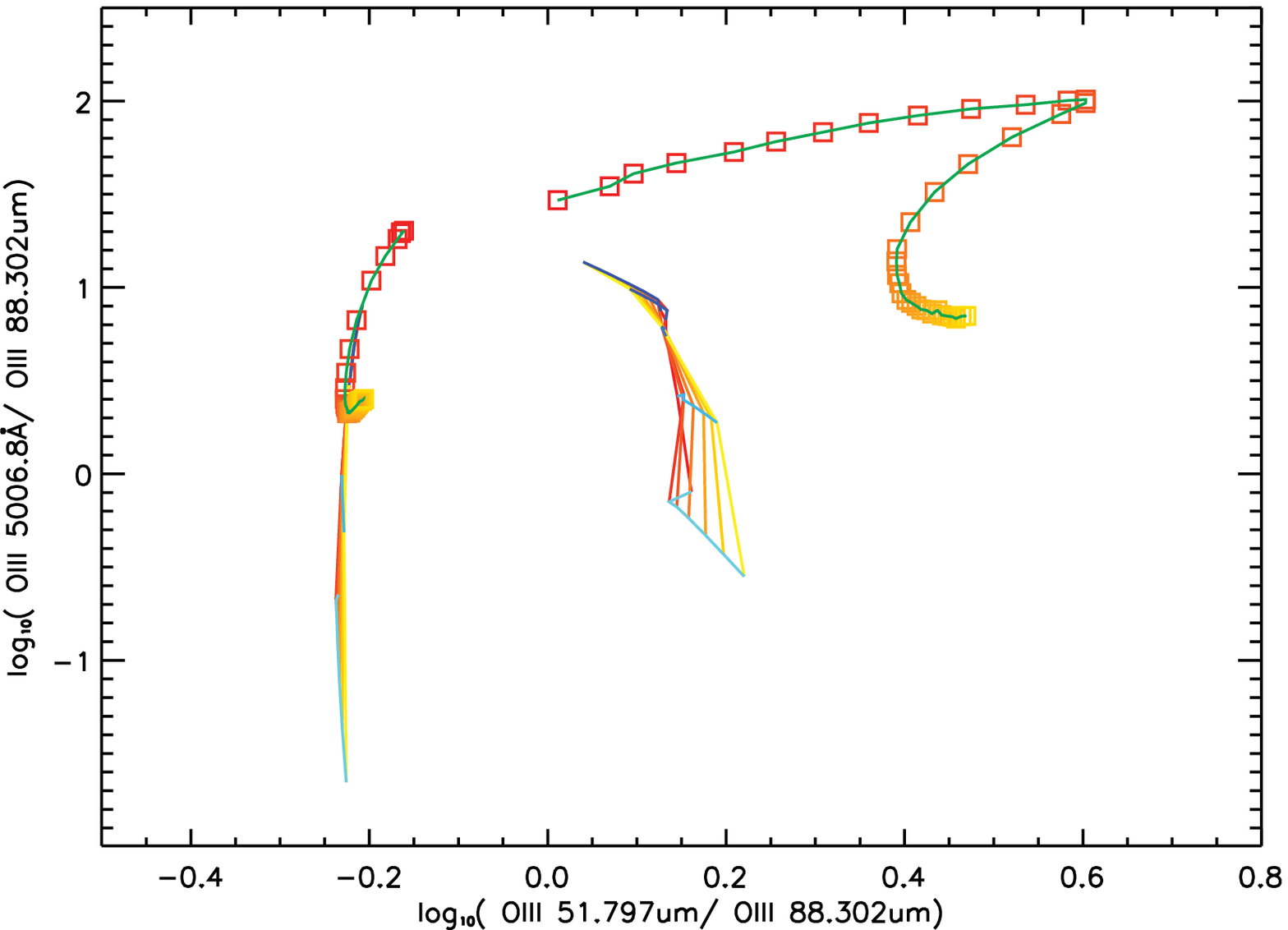}
\caption{Diagnostic diagram using only O$^{+2}$ emission lines,
  containing the density sensitive [\oiii] $52\mum/88\mum$ ratio
  versus the temperature sensitive [\oiii]$5007$\AA$/88\mum$
  ratio, that both use the \emph{Herschel} observable [\oiii]$88\mum$
  line. 
  Shown are two different density ($n_{\rm H}=10\pccm$, lower left
  grid, and $n_{\rm H}=350\pccm$, lower right grid) instantaneous
  starburst models from \citet{Kewley01}, and two different density
  ($n_{\rm H}=0.1\pccm$, upper left green curve, and $n_{\rm
    H}=10\pccm$, upper right green curve) shock models from
  \citet{Allen08}. In both models, the state of ionization (ionization
parameter or shock velocity) increases from red to yellow. The four
models are well separated in this diagram thanks to the diagnostic
power of the ratios.}
\label{fig:oiii}
\end{figure}

The final diagram uses optical and far-IR lines of the same species,
O$^{+2}$, aiming towards the  ionized gas diagnostics possible with
\emph{Herschel Space Telescope}, specifically the PACS spectrometer. Figure
\ref{fig:oiii} shows the [\oiii] $52\mum/88\mum$ versus
[\oiii]$5007$\AA$/88\mum$, based on the \citet{Dinerstein85}
diagram. This diagram is a strong diagnostic diagram, separating out
both density and temperature, while avoiding direct abundance and
ionization effects. While unfortunately the density sensitive
$52\mum/88\mum$ ratio cannot be observed in the local Universe with
\emph{Herschel} due to the lower wavelength limit, the temperature
sensitive $5007$\AA$/88\mum$ ratio can be obtained with the
combination of \emph{Herschel} PACS spectra with
ground based data. In reality however, caution must be taken with this
ratio, as there are clear difficulties associated with the different
wavelengths, such as matching the \emph{Herschel} apertures and pixels
with ground based data, and the large difference in dust attenuation
between the two lines, both of which effects could overwhelm any small
variations due to temperature.
Nevertheless, to demonstrate the power of this diagram we show
two star forming model grids from \citet{Kewley01}, and two shock model
curves from \citet{Allen08}.  

The \citet{Kewley01} models are grids in
metallicity (blue curves, same as Levesque models) and ionization
parameter (orange curves) using a  Starburst99 3 Myr old instantaneous
burst as the ionizing source. What distinguishes the two model grids
is the density of the ionized gas, with the left grid at $n_{\rm H}=10\pccm$, and the
right grid at $n_{\rm H}=350\pccm$. The $52\mum/88\mum$ ratio clearly
separates the two grids. Though difficult to see due to their
collapsed form, the $5007$\AA$/88\mum$ ratio is affected by both the
ionization parameter (weakly) and the metallicity (more
strongly). However the diagnostic strength of this ratio is more
clearly seen when compared to the much hotter ionized gas in the
shocks. 

The \citet{Allen08} model curves (green) show pure shock models at solar
metallicity, magnetic parameter of   B/$\sqrt{\rm n}=1\mu$G\
cm$^{3/2}$, and velocities ranging from 150 (red squares) -- 1000
(yellow squares) \kms. Note that the velocities are shown as squares
due to one of the features of ITERA, where if one of the parameters
are set to a single value, in this case solar metallicity, the varying
parameter along the curve is marked by squares. As with the Kewley
models, the two curves are again separated by density,
with the left curve having a pre-shock density of $n_{\rm H}=0.1\pccm$, and
the right $n_{\rm H}=10\pccm$. What is clearly noticeable in figure
\ref{fig:oiii}, is the clear vertical offset between the shock models
and star forming models in the temperature sensitive
[\oiii] $5007$\AA$/88\mum$ ratio, demonstrating clearly the possible
capabilities of upcoming \emph{Herschel}-PACS spectral surveys.

\section{Summary}

We have presented here a new tool to enable astronomers to compare
observations of emission line ratios with that determined by
photoionization and shock models, ITERA, the IDL Tool for
Emission-line Ratio Analysis. ITERA is an IDL widget tool which allows
the user to plot ratios of any strong atomic and ionized emission lines as
determined by standard photoionization and shock models.  These line
ratio diagrams can then be used to determine diagnostics for nebulae
excitation mechanisms \citep[such as discussed in
the][paper]{Baldwin81} or for nebulae parameters such as density,
temperature, metallicity, etc. ITERA can also be used to determine
line sensitivities to such parameters, compare observations with the
models, or even estimate unobserved line fluxes. This tool, and
associated libraries and instructions, can be obtained at
\url{http://www.brentgroves.net/itera.html}. 

At its core lies a library of emission-line nebulae models, covering
photoionization by young O and B stars
\citep[e.g.][]{Kewley01,Dopita06}, photoionization by a hard power-law
spectrum from an AGN \citep[e.g.][]{Groves04b} and ionization by shocks
\citep[e.g.][]{Allen08}. These models broadly cover the space of
observed emission-line objects.

 Finally, to demonstrate the capabilities of ITERA we present three
 different line diagnostic diagrams; the classic BPT diagram of
 [\nii]$\lambda\lambda6548,6584/ha$ versus [\oiii]$\lambda5007/\hb$,
 showing the coverage of the star forming galaxy and AGN
 photoionization models when compared to SDSS emission-line galaxies,
 the mid-IR diagram of 
 [\neiii]${15.6\mu\rm{m}}$/[\neii]${12.8\mu\rm{m}}$ versus
 [\siv]${10.5\mu\rm{m}}$/[\neii]${12.8\mu\rm{m}}$, revealing an
 interesting correlation of ionization parameter and metallicity of
 stellar photoionization models when compared to observations, and the
 optical IR diagram of O$^{+2}$,  [\oiii] $52\mum/88\mum$ versus
[\oiii]$5007$\AA$/88\mum$, demonstrating one of the diagnostic possibilities of
\emph{Herschel}. 

It is hoped that this tool will be of great use to the astronomical
community when analyzing spectra of emission-line regions due to both
its simplicity and ease of use, especially as the library expands with
the growing number of photoionization and shock models becoming
available. 

\section*{Acknowledgements}{The authors would like to thank those who have given
feedback on ITERA as it was being developed; B. Brandl, E. da Cunha,
J. Holt, M. Sarzi and H. Spoon.}

\end{document}